\documentclass[11pt]{article}
\usepackage[a4paper,left=2.85cm,right=2.85cm,top=2.5cm,bottom=2.5cm]{geometry}
\usepackage[colorlinks,allcolors=blue]{hyperref}
\newcommand{\articletype}[1]{}
\newcommand{\orcid}[1]{}
\providecommand{\keywords}[1]{\par\medskip\noindent\textbf{Keywords:} #1\par}
\providecommand{\roles}[1]{\section*{Author contributions}#1}
\providecommand{\ack}[1]{\section*{Acknowledgments}#1}
\providecommand{\funding}[1]{\section*{Funding}#1}
\providecommand{\data}[1]{\section*{Data availability}#1}
\usepackage{amsmath,amssymb}
\usepackage{lmodern}
\usepackage{graphicx}
\usepackage{bm}
\usepackage{booktabs}
\usepackage[numbers,sort&compress]{natbib}
\usepackage{microtype}
\usepackage{url}

\makeatletter
\let\tableinput\@@input
\makeatother
\begin{document}

\title{Structural identifiability and stress reconstruction from incomplete optical maps with velocimetry}
\author{Zijian Liu$^{1,*}$, Julian Olszewski$^{1,*}$, Bruce I. Gaynes$^{2}$ and Jie Xu$^{1,\dagger}$}
\date{}
\maketitle

\begin{center}\small
$^{1}$ Department of Mechanical and Industrial Engineering, University of Illinois Chicago, Chicago, IL 60607, USA\\
$^{2}$ Department of Ophthalmology, Stritch School of Medicine, Loyola University Chicago, Maywood, IL 60153, USA\\
$^{*}$ These authors contributed equally to this work as co-first authors.\\
$^{\dagger}$ Corresponding author: jiexu@uic.edu
\end{center}

\begin{abstract}
Reconstructing the stress field of a planar viscoelastic flow from optical
measurements loses its direct evidence wherever optical coverage is interrupted, and
no improvement in optical precision restores an observation that was never made. We characterize what a second, velocity channel adds, and what
neither channel can supply. Two calibrated optical components determine the local
deviatoric stress pointwise, while velocity constrains spatial stress variation through
momentum balance, so the two channels are complementary rather than redundant. The
isotropic part of the stress is unobservable to both: the divergence of an isotropic field
is a pure gradient, which the Leray projection annihilates, so every representable
isotropic mode lies in the joint null space. That accounts for the null space exactly
when the optical field is complete, and bounds it from below otherwise, since finite
incomplete sampling and aperture zeros can remove further directions. We verify the
count directly on three discretizations. In paired synthetic tests with finite measurement apertures,
spatially correlated noise and optical stripe dropout, adding velocity reduces the mean
whole-domain deviatoric error from $50.40\%$ to $27.82\%$ at $3\%$ reference noise, and
the improvement survives shared gaps, inverse-grid refinement at fixed physical sampling,
and a constitutively generated stress field. The improvement does not rest on how the
regularization parameter is chosen: it holds under both the expected-norm discrepancy
rule and generalized cross-validation, and we report each selection with its position
in the search interval, which is where the two rules differ.
\end{abstract}


\section{Introduction}\label{sec:intro}
When a dilute polymer solution flows, the dissolved chains stretch and align.
The elastic force they carry enters the momentum balance as an extra stress: a
symmetric tensor defined at every point of the flow. That stress field sets the
traction on the walls and the force on suspended particles, and it is the quantity a
constitutive model has to predict; none of it follows from the velocity field alone.
Reading it calls for signals gathered through a window from outside, because a probe
placed in the channel disturbs the flow it is meant to characterize.

Two such signals are standard in this setting, and they exploit different physics. The
first is optical. A stretched polymer solution becomes birefringent, meaning its
refractive index depends on direction, so polarized light crossing the channel emerges
with a phase difference between two polarization components (the retardance) and
with a characteristic orientation (the azimuth). Under a calibrated planar stress-optic relation
those two numbers are set by the stress where the light passed, and together they fix
the local deviatoric stress
pointwise~\cite{maxwell1874,lodge1956,philippoff1961,fuller,janeschitz1983}. The second
signal is velocimetric: tracer particles are seeded in the fluid and imaged in quick
succession, and correlating successive images returns the velocity
field~\cite{haward2013,vanoudheusden2013}. Velocity does
not report stress at a point. It is tied to stress through the force balance, which
involves the divergence of the stress, so it constrains how stress varies from place to
place rather than what it is anywhere in particular.

That difference is what makes the combination worth analyzing, and it is clearest in
terms of the standard splitting of a stress tensor. Any symmetric tensor is the sum of
an isotropic part, one number per point that acts equally in all directions much as a
pressure does, and a deviatoric part, which carries the directional differences. Which
of the two a channel can see turns out to decide what the pair can deliver, and we
return to it below.

The practical obstacle is that optical coverage is often incomplete. Windows are obscured,
processed correlation maps are discarded where they fail quality gates, and finite
averaging apertures blur what remains~\cite{haward2013,nekkanti2023}. A gap of this kind
is not a precision problem: no improvement in optical signal-to-noise supplies a
direct observation where none was made, and whatever is recovered there comes from the
representation, a prior, or another channel.

Because momentum balance couples stress at one location to stress in its
neighborhood, a velocity field carries information about stress where the optical map
is blank. That raises two questions that are structural rather than statistical. How much does the velocity channel actually
recover across a gap, and what can neither channel supply at all? Both are questions about
the null space of the joint observation operator, and neither is answered by reporting a
noise level.

The second question has a clean answer that organizes the rest of this paper. The
isotropic part of the stress is invisible to \emph{both} channels. Two optical components
see only the deviatoric part by construction. Velocity sees the divergence of the stress,
and the divergence of an isotropic field is a pure gradient, which the Leray projection
that defines the velocity response annihilates. The joint operator is therefore rank deficient by at least
the number of representable isotropic modes for any sampling, and by exactly that
number when the optical field is complete. Finite incomplete sampling and aperture
zeros can remove further directions, so the gauge is a floor on what no measurement
of this kind supplies rather than the whole of it. The reconstruction target must in
either case be stated as a quotient that excludes the gauge. We verify the count
numerically on three discretizations rather than assuming it.

Using mechanical equilibrium to propagate stress across a field where measurement is
incomplete is a classical idea in photoelasticity, where stress-separation methods
integrate equilibrium from a known boundary~\cite{frocht1941,ramesh2000,solaguren2010};
the velocity channel used here is a field-level analogue that requires no boundary
integration. In the viscoelastic setting, stress inference more often embeds the governing
equations directly in the fit~\cite{raissi2019,raissi2020,thakur2024,nnpinn2022,xu_tartakovsky2021},
or follows fluid trajectories and integrates a specified constitutive
law~\cite{kumar2023,majidi2026}. Combined optical and velocity measurements in planar
contractions and microfluidic rheo-optics provide the experimental settings in which the
complementarity of the two channels can be
studied~\cite{quinzani1994,ober2011,salipante2025}. The structural setting for
divergence-free symmetric tensor fields, which is what the velocity channel cannot see, is
the differential complex of continuum mechanics~\cite{angoshtari}.

This paper treats the combination as an inverse problem. We identify the stress
information supplied by each channel, establish the gauge that neither supplies, and
quantify reconstruction from incomplete optical maps in paired synthetic tests with finite
apertures and spatially correlated noise. We then delimit the scope of the regularization selection, which the agreement of
two standard rules does not by itself establish. The two kinds of result carry different weight. The identifiability statements
follow from the observation operators themselves, so they hold for any sampling of
this measurement pair; the reconstruction figures are properties of the stated
ensembles, apertures and covariances.

Section~\ref{sec:models} establishes the observation operators, the gauge and the physical
error metric. Section~\ref{sec:reconstruction} quantifies reconstruction from incomplete
optical maps, including fixed-aperture refinement and a constitutively generated stress
field. Section~\ref{sec:discussion} delimits the scope of the regularization selection.
Derivations and numerical sensitivity analyses are given in the Supplementary Material.

\section{Measurement structure and reconstruction metric}\label{sec:models}

\subsection{Velocity and calibrated stress observations}

Let $\bm\sigma$ denote the symmetric polymer extra-stress. In planar steady
creeping flow, momentum and incompressibility give
\begin{equation}
-\nabla p+\eta_s\Delta\bm u+\nabla\cdot\bm\sigma=\bm0,
\qquad \nabla\cdot\bm u=0.
\label{eq:stokes}
\end{equation}
On an endpoint-free periodic domain with fixed mean velocity, the
pressure-ambiguous velocity map is
\begin{equation}
\mathcal V\bm\sigma=(-\eta_s\Delta)^{-1}
\mathcal P(\nabla\cdot\bm\sigma),
\label{eq:velop}
\end{equation}
where $\mathcal P$ removes gradient forcing. Particle image velocimetry (PIV)
supplies spatial averages of this velocity field in the reconstruction
experiment. An ideal
force-referred observation instead measures
$\mathcal D\bm\sigma=\nabla\cdot\bm\sigma$ after pressure and solvent
contributions have been specified. PIV-based pressure recovery under a known Newtonian stress~\cite{vanoudheusden2013}
is an instance of supplying that additional modeling information.

The calibrated optical coordinates are
\begin{equation}
\mathcal B\bm\sigma=(b_1,b_2)
=(N_1,2\sigma_{xy}),\qquad N_1=\sigma_{xx}-\sigma_{yy}.
\label{eq:pair}
\end{equation}
For a spatially resolved optical field, they determine the planar
deviatoric stress pointwise:
\begin{equation}
\operatorname{dev}\bm\sigma=\mathcal R\bm b
=\frac12\begin{pmatrix}b_1&b_2\\b_2&-b_1\end{pmatrix}.
\label{eq:dense_inverse}
\end{equation}
Equation~\eqref{eq:dense_inverse} is both the direct optical estimator and
the baseline for assessing an additional velocity measurement. Applying
the same spatial average to both optical components gives the corresponding
averaged deviatoric stress. The numerical observation matrices include each
sensor's spatial response explicitly. Throughout the numerical studies, the term optical data refers to these
calibrated polymer-stress coordinates. The material and optical forward calibration is set out in the companion
measurement-design paper~\cite{companion}.

\subsection{Two-channel stress closure}

A divergence-free symmetric stress leaves momentum unchanged. Locally, its
Airy representation is
\begin{equation}
\mathcal A\phi=
\begin{pmatrix}\phi_{,yy}&-\phi_{,xy}\\-\phi_{,xy}&\phi_{,xx}\end{pmatrix},
\qquad \nabla\cdot\mathcal A\phi=\bm0.
\label{eq:airy}
\end{equation}
On a simply connected planar domain, a smooth symmetric divergence-free
stress admits a global Airy representation~\cite{angoshtari}.
On a periodic torus, nonzero Fourier modes have this representation and
constant symmetric stresses supply the remaining modes. The Supplementary Material treats topology and discrete
compatibility explicitly.

\begin{figure}[tbp]
\centering
\includegraphics[width=0.98\textwidth]{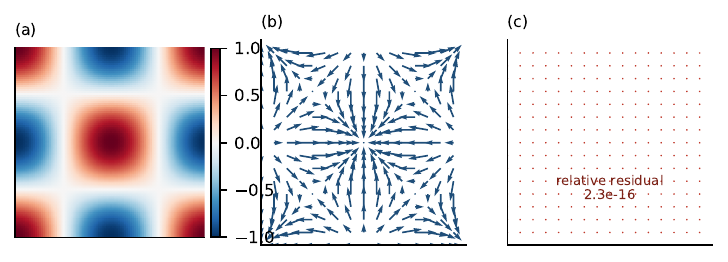}
\caption{\label{fig:gauge}
The gauge that neither channel observes, illustrated on the periodic torus. (a) An isotropic stress mode $\sigma=p\,I$ contributes nothing to the optical pair, which measures only deviatoric components. (b) Its divergence is the pure gradient $\nabla p$. (c) The Leray projection that defines the velocity response annihilates any gradient, so the velocity channel does not see it either; the residual shown is at machine precision. The joint operator is therefore rank deficient by at least the number of representable isotropic modes however densely either modality is sampled, and the reconstruction metric must quotient this component out.
}
\end{figure}

For pressure-ambiguous velocity, adding $q(x,y)\bm I$ to the stress is
absorbed by pressure. Since
$\bm\sigma=\mathcal R\mathcal B\bm\sigma+
\tfrac12\operatorname{tr}(\bm\sigma)\bm I$, the periodic map satisfies
\begin{equation}
\mathcal V=\mathcal V\mathcal R\mathcal B,
\qquad \ker[\mathcal V;\mathcal B]=\ker\mathcal B
=\{q(x,y)\bm I\}.
\label{eq:dense_factorization}
\end{equation}
A spatially complete calibrated optical field therefore supplies the
full noiseless deviatoric field. Additional velocity observations can still help where noise averaging,
regularization or incomplete spatial sampling limit the estimate. For the force-referred map,
joint divergence and optical data additionally constrain $\nabla q$, leaving
only a constant isotropic stress on a connected domain.
Figure~\ref{fig:gauge} shows the mode that neither channel returns.

This closure identifies the role of mixed measurements. Optical observations
resolve the local deviatoric components, while velocity couples stress
variations across the domain through momentum balance. At incomplete spatial
coverage, that coupling supplies constraints between optical sites. The
reconstruction study measures this contribution with finite apertures and
incomplete optical coverage.

\subsection{Physical reconstruction metric}

We measure stress using tensor Frobenius energy,
\begin{equation}
\|\bm\sigma\|_F^2=\int_\Omega
(\sigma_{xx}^2+\sigma_{yy}^2+2\sigma_{xy}^2)\,d\bm x,
\label{eq:frobenius}
\end{equation}
so that
\begin{equation}
\|\operatorname{dev}\bm\sigma\|_F^2
=\frac12\int_\Omega [N_1^2+(2\sigma_{xy})^2]\,d\bm x.
\label{eq:devidentity}
\end{equation}
The reconstruction error is
\begin{equation}
e_{\rm dev}=
\frac{\|\operatorname{dev}(\widehat{\bm\sigma}-\bm\sigma^\star)\|_F}
{\|\operatorname{dev}\bm\sigma^\star\|_F}.
\label{eq:quotienterror}
\end{equation}
This metric compares the observable planar stress while removing the exact
isotropic gauge. With a coordinate change
$T=\operatorname{diag}(1,2)$ from $(N_1,\sigma_{xy})$, noise covariance
transforms as $\Sigma_{\rm phys}=T\Sigma_{\rm legacy}T^{\mathsf T}$.
Consistent whitening preserves the information under this change of units.

\subsection{Domains and decision criteria}

The reconstruction study uses an endpoint-free torus, fixed physical
apertures and uniform physical mass, and it states its observation operator,
covariance and decision criterion explicitly. The Supplementary
Material analyzes finite-window sampling and explicit boundary information
as separate observation settings, using trapezoidal window mass and
pressure elimination against the full discrete gradient range.

The reconstruction comparison fixes observation operators, methods and
paired noise draws before evaluating performance, so the reported error
differences isolate the contribution of the added velocity channel.

\paragraph{Numerical implementation}
Numerical rank counts singular values above
$\tau=\max(m,n)\epsilon_{\rm mach}\sigma_{\max}$ for an $m\times n$
matrix. The regularized reconstruction retains all singular values;
this threshold is used only for rank diagnostics. Relative-threshold
sensitivity is a separate conditioning diagnostic.
The Supplementary Material gives the absolute-threshold and finite-field
rank witnesses. All figures are generated from recorded numerical outputs, and the environment,
seeds, covariance, regularization choices and residual checks are stated in the
Supplementary Material.

\section{Stress reconstruction from incomplete optical maps}\label{sec:reconstruction}

\subsection{Paired finite-resolution measurements}

Spatial averaging and measurement coverage determine how local optical stress
information combines with the spatial coupling supplied by velocity. We study
this interaction using 40 paired synthetic stress fields on the periodic
domain $[-1,1)^2$, with $\eta_s=1$ in Eq.~\eqref{eq:velop}. The fields are
smooth trigonometric polynomials with component wavenumbers up to 8,
spectral length $0.24$, spectral power 3, and imposed Airy energy fraction
$0.35$. This ensemble isolates measurement effects from constitutive fitting.
The optical coordinates and reconstruction error follow
Eqs.~\eqref{eq:pair} and~\eqref{eq:quotienterror}.

The candidate optical map has 289 centers on a uniform $17\times17$
lattice. PIV uses the 81 stride-2 centers of that lattice. Each observation
is an exact periodic square average, with fixed physical side length
$2/17$ for optics and $8/17$ for PIV. These widths represent spatial
binning and interrogation-window averaging, as used in rheo-optical and
velocimetry measurements~\cite{haward2013,salipante2025}. The selected
dimensionless widths define an instrument-informed synthetic comparison.
The inverse representation is independent of the sensor lattice: the primary
reconstruction uses $33\times33$ Fourier modes, and refinement retains the
same continuous truths, physical centers, apertures and measured values.

We compare a complete optical output map with two geometric patterns of
missing outputs: a stripe $|x|<0.30$ and a disk $x^2+y^2<0.60^2$.
These remove 102 and 80 optical centers, respectively. A shared-stripe
control also removes 27 PIV centers. Each mask selects available processed
outputs after spatial averaging; retained apertures preserve their full
spatial response. This models incomplete measurement maps rather than
changing the fluid domain. Full, stripe, disk and shared-stripe coverage therefore
retain $289$, $187$, $209$ and $187$ optical centers; the first three retain all $81$
velocimetric sites and the shared stripe retains $54$.

Each realization supplies one full noise map per modality, shared by all
its optical-only, PIV-only and joint reconstructions. The standard deviations
are $3\%$ of fixed reference RMS scales for velocity and raw optical
coordinates $(N_1,\sigma_{xy})$. Within-site correlations are $+0.20$ and
$-0.25$, respectively. Optical noise and covariance are both transformed by
$T=\operatorname{diag}(1,2)$. The primary spatial covariance has a periodic
Gaussian component of weight $0.4$ and length $0.20$, plus an independent
nugget; the two modalities have independent noise. For each observed subset,
we select its covariance principal submatrix and recompute the Cholesky
whitener. The Supplementary Material specifies the kernel and reference
scales and compares spatially independent noise and point observations.

In Fourier coordinates orthonormal for area-mean Frobenius energy, let $O$
be the selected aperture-averaged observation matrix, $W$ its whitener,
and $R_\ell$ the inverse factor of the coercive spectral $H^1$ penalty.
We reconstruct $R_\ell w$ by minimizing
\begin{equation}
\|WOR_\ell w-Wy\|_2^2+\alpha\|w\|_2^2,
\label{eq:paired_tikhonov}
\end{equation}
with $\ell=0.15$. Expected-norm Morozov selects $\alpha$ to give whitened
residual norm $\sqrt m$, where $m$ is the scalar observation count.
Generalized cross-validation (GCV) supplies a second selection rule. It picks
$\alpha$ to minimize an estimate of the error in predicting a withheld
measurement, so it requires no declared noise level.
The search interval spans $10^{-14}$ to $10^6$ times
$\sigma_{\max}(WOR_\ell)^2$, with 121 logarithmic GCV candidates.
Both rules use the observations and specified noise covariance.
Two-sided $95\%$ Student-$t$ intervals use the 40 within-realization error
differences. These pointwise intervals describe the prescribed synthetic
ensemble; the stripe comparison with Morozov is the primary contrast.

\subsection{Velocity information in incomplete optical maps}
\begin{figure}[tbp]
\centering
\includegraphics[width=\textwidth]{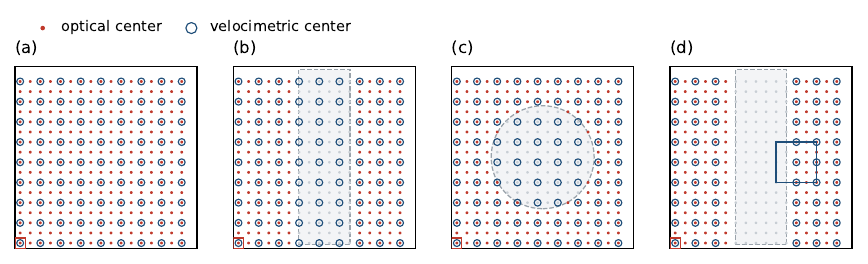}
\caption{\label{fig:coverage}
The four output-map settings, on the $17\times17$ candidate grid. (a) Full, (b) stripe,
(c) disk and (d) shared stripe retain 289, 187, 209 and 187 optical centers; the first
three retain all 81 velocimetric centers and the shared stripe retains 54. Shading marks
the rejected region, $|x|<0.30$ for the two stripes and $|\bm x|<0.60$ for the disk.
Small filled markers are retained optical centers, pale markers are rejected ones, and
open markers are velocimetric centers. The squares are measurement apertures drawn to
scale, $2/17$ for optics and $8/17$ for velocimetry. The large square in (d) is the
retained velocimetric aperture nearest the gap: masks reject output centers after
averaging, so an aperture beside the gap still overlaps it, and the shared-stripe control
therefore removes velocimetric outputs from the gap rather than making the velocity
channel blind there.
}
\end{figure}

\begin{figure}[tbp]
\centering
\includegraphics[width=\textwidth]{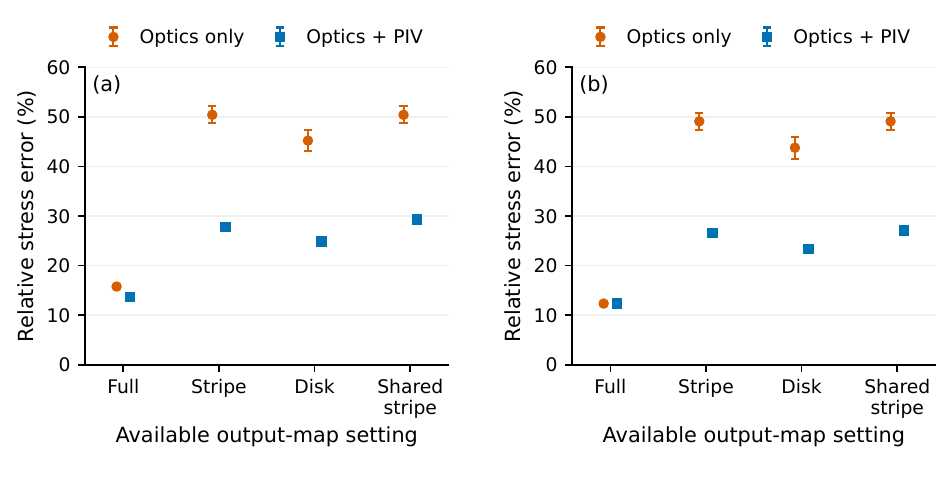}
\caption{\label{fig:imaging}
Stress reconstruction from incomplete optical maps.
Mean whole-domain deviatoric Frobenius error for optical-only and joint
optical--PIV reconstruction using (a) expected-norm Morozov and (b) GCV.
The reconstruction grid is $33\times33$, with fixed square-aperture widths
$2/17$ for optics and $8/17$ for PIV, $3\%$ reference noise, and the
specified spatially correlated covariance. Full, stripe, disk and shared-stripe
settings retain 289, 187, 209 and 187 optical centers. Joint measurements
retain 81 PIV sites except for the shared stripe, which retains 54.
Masks select available output centers after averaging. Error bars are
pointwise $95\%$ confidence intervals for each arm's mean across
40 paired synthetic realizations.}
\end{figure}

Figure~\ref{fig:coverage} shows the four output-map settings and the
apertures that serve them. Joint measurements improve stress reconstruction in every
missing-map case (Figure~\ref{fig:imaging} and Table~\ref{tab:regional}). Two
features of Table~\ref{tab:regional} matter more than the headline number. The improvement
is concentrated where it should be: inside the gap, where optical-only reconstruction has its
largest error, the joint estimate roughly halves it, while in the observed
region, where optics is already good, it changes little. The two selection rules also agree throughout these tests, which is an observation
rather than a guarantee; section~\ref{sec:discussion} gives the reason the
construction does not predict it. The joint whole-domain estimate also improves on
the matched PIV-only error of $46.58\%$, so the gain is not simply the velocity channel
acting alone.

\begin{figure}[tbp]
\centering
\includegraphics[width=\textwidth]{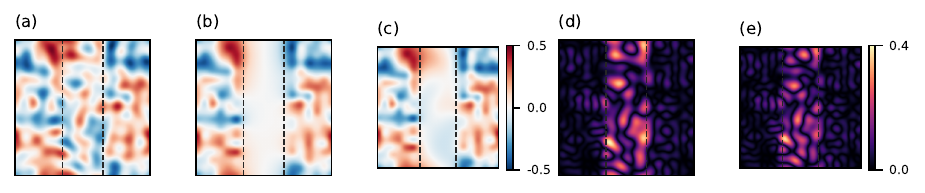}
\caption{\label{fig:field}
What the velocity channel recovers inside the gap, for the first realization of the
primary stripe case. The realization index is fixed in advance rather than selected for
appearance; its whole-domain deviatoric errors are $39.60\%$ for optics alone and
$27.11\%$ jointly, against ensemble means of $50.40\%$ and $27.82\%$. Panels show
the normal-difference coefficient $a=N_1/\sqrt2$ of the true stress (a), the
optical-only reconstruction
(b), the joint reconstruction (c), and the absolute errors of those two reconstructions
(d) and (e). Panels (a) to (c) share one symmetric color scale and panels (d) and (e)
share a second; dashed lines mark the dropped stripe $|x|<0.30$. Reconstruction uses
expected-norm Morozov on the $33\times33$ inverse grid with the same fixed apertures,
spatially correlated covariance and $3\%$ reference noise as
Figure~\ref{fig:imaging}.
}
\end{figure}

Figure~\ref{fig:field} shows one realization of the stripe case as fields:
the optical-only estimate loses the structure inside the dropped stripe, and the joint
estimate recovers it. The disk and shared-stripe controls in Figure~\ref{fig:imaging}
show the same complementary use of the two channels. The shared-stripe case is the informative one: when
both modalities lose their output centers over the same region, the joint estimate
still improves on optics alone. The masks reject output centers after averaging, so
retained apertures next to the stripe still reach into it; the control shows that the
gain survives removing velocity outputs from the gap, not that the velocity channel is
blind there.

Complete optical coverage gives the reference point for interpreting the
gain. With the same apertures and correlated noise, optics alone reaches $15.78\%$ and the joint estimate $13.70\%$ (Figure~\ref{fig:imaging}), so with full coverage the velocity
channel adds little: its contribution is specific to incomplete maps rather than a general
improvement. Figure~\ref{fig:imaging} collects the four coverage settings under
both selection rules.

\begin{table}[tbp]
\centering
\caption{\label{tab:regional}
Stripe-dropout reconstruction by region and selection rule. Mean deviatoric error over 40
paired realizations, $3\%$ spatially correlated noise, $n=33$ inverse grid. Regional errors
are normalized by the true stress energy of their own region. The paired whole-domain
decrease under Morozov is $22.58$ percentage points, with pointwise $95\%$ interval
$[20.95,24.22]$.}
\small
\begin{tabular}{lrrrr}
\toprule
& \multicolumn{2}{c}{Morozov} & \multicolumn{2}{c}{GCV} \\
\cmidrule(lr){2-3}\cmidrule(lr){4-5}
Region & Optics (\%) & Joint (\%) & Optics (\%) & Joint (\%) \\
\midrule
Whole domain      & 50.40 & 27.82 & 49.09 & 26.55 \\
Inside the gap    & 86.34 & 44.29 & --    & --    \\
Observed region   & 21.86 & 17.03 & --    & --    \\
\bottomrule
\end{tabular}
\end{table}

\subsection{Fixed-aperture refinement}

Refining the inverse representation to $49\times49$ and $65\times65$ while holding
the physical measurements fixed changes the stripe results by less than $0.07$
percentage points. Across the full
and stripe maps, both selectors and all three modality configurations, final-step mean
error changes stay below $0.071$ percentage points and mean field changes below $0.70\%$ of
the true-field norm. The reconstruction is therefore stable over the tested refinement at
fixed bandlimited truth and nonzero physical aperture, which is the property that lets the
reported errors be read as measurement-limited rather than discretization-limited.

Both experiments retain the complete singular spectrum in the regularized
solve. Their normalized normal-equation residuals remain below
$1.1\times10^{-14}$.
The Supplementary Material records regularization choices, GCV boundary
selections, covariance checks and the distinction between new solves and
repeated output records.

\subsection{Constitutively generated stress and sensitivity}

A four-roll finitely extensible nonlinear elastic Peterlin (FENE-P) stress field
provides a complementary reconstruction example. The Peterlin closure imposes a finite maximum chain
extension through a nonlinear spring force that increases as the conformation
approaches that limit. We solve the equilibrium-normalized three-dimensional constitutive
model with dynamic $A_{zz}$ in a prescribed periodic four-roll flow, using
$\mathrm{Wi}=0.8$, $L^2=50$ and conformation diffusivity $0.002$.
The observation velocity is the stress-induced Stokes response
$\mathcal V\bm\sigma$. A known-force momentum completion relates this
response to the prescribed flow. The source uses $95\times95$ points,
and its full Fourier bandwidth is retained in the observations and error.

Because that completing force is obtained from the generated stress rather
than specified independently, it is worth stating what it does and does not
affect. The Supplementary Material gives it explicitly as
$\bm f=-\eta\Delta\bm u_{\rm prescribed}-\nabla\cdot\bm\sigma$ with $p=0$,
and records that the observation velocity is distinct from the prescribed
four-roll velocity used to generate the conformation field. The force enters
only the \emph{generation} of the example. It does not appear in the
observation operator: by Eq.~\eqref{eq:velop} the observable is
$\mathcal V\bm\sigma$, a function of the stress alone, and $\mathcal V$ is the
same operator used in the inversion. No reconstruction result here therefore
depends on the completing force, and none of the reported errors would change
if it were specified differently while $\bm\sigma$ was held fixed.

The consequence is a limit on interpretation rather than on validity. This
example tests the observation operators against a stress field produced by an
actual constitutive model, with the spatial structure and finite-extensibility
saturation that entails; it is not a demonstration that flow, stress and
momentum balance are mutually self-consistent. The channel study of
the companion measurement-design paper~\cite{companion} is where that self-consistency is assessed. We
therefore read the four-roll numbers as constitutive realism for the
reconstruction operator, and the channel numbers as the self-consistent
case.
Forty independent noise realizations use $3\%$ of the aperture-averaged
response RMS at the 81 PIV and 289 optical centers. This common covariance
is held fixed across the constitutive-field comparisons.

Table~\ref{tab:fourroll} collects the results. The pattern of
Table~\ref{tab:regional} repeats on a stress field produced by integrating a
constitutive model rather than drawn from the reconstruction ensemble, with the
spatial structure and finite-extensibility saturation that entails. The joint
estimate again improves substantially under stripe dropout and barely at all under
complete coverage.
Refining the inverse grid to $65\times65$ changes the joint stripe error only to
$41.42\%$, and the missing-map improvement persists across stripe half-widths
$0.15$--$0.45$ and prior lengths $0.075$--$0.30$ in the prespecified comparisons.
The largest mean field change across the tested inverse-grid pairs and selectors
is below $0.85\%$ of the source-stress norm.

The source-grid comparison, from $63\times63$ to $95\times95$, changes
the generated deviatoric field by $5.41\%$ of the finer-field norm.
This source-resolution sensitivity is reported separately from fixed-truth
inverse refinement. The Supplementary Material gives every arm, regional
error and paired interval, together with the distinct response-relative
noise calibration. It also varies the random ensemble's full-tensor Airy
fraction from $0.15$ to $0.85$: optical and joint Morozov errors are
$48.33\%$ and $18.66\%$ at the lower value, and $58.30\%$ and
$49.84\%$ at the upper value. These comparisons locate the gain within
the stress content, measurement coverage and prior used for reconstruction.

\begin{table}[tbp]
\centering
\caption{\label{tab:fourroll}
Reconstruction of a constitutively generated four-roll FENE-P stress field at
$\mathrm{Wi}=0.8$, $L^2=50$. Mean deviatoric error over 40 realizations at $3\%$ of the
aperture-averaged response RMS, $n=33$ inverse grid unless stated. Errors inside the gap
are normalized by the true stress energy of that region.}
\small
\begin{tabular}{llrrrr}
\toprule
& & \multicolumn{2}{c}{Morozov} & \multicolumn{2}{c}{GCV} \\
\cmidrule(lr){3-4}\cmidrule(lr){5-6}
Coverage & Region & Optics (\%) & Joint (\%) & Optics (\%) & Joint (\%) \\
\midrule
Stripe dropout & whole domain   & 65.10 & 41.39 & 64.99 & 40.11 \\
Stripe dropout & inside the gap & 96.71 & 61.01 & --    & --    \\
Complete       & whole domain   & 14.07 & 13.94 & --    & --    \\
\bottomrule
\end{tabular}
\end{table}

\section{Scope of the regularization selection}\label{sec:discussion}
\begin{figure}[tbp]
\centering
\includegraphics[width=0.86\textwidth]{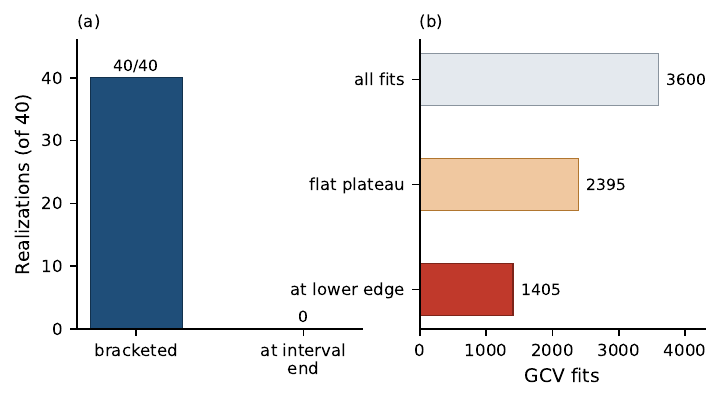}
\caption{\label{fig:selector}
Recorded regularization selections. (a) Every expected-norm Morozov target is bracketed within the search interval. (b) Among the $3\,600$ distinct GCV fits in the equal-count and imaging studies, $1\,405$ select the lower grid edge and none the upper; $2\,395$ meet the near-flat small-parameter plateau diagnostic. All selected fits are included in the reported reconstruction errors. The Supplementary Material defines the diagnostics and gives the per-fit records.
}
\end{figure}

The stripe-map improvement holds under both expected-norm Morozov and GCV, although
the two rules select different parts of the search interval. Figure~\ref{fig:selector}
summarizes that distinction; its interpretation depends on the attainable residual
range and the shape of the selection criterion.

For a whitened operator $A$ and
data $y$, the Tikhonov residual increases monotonically from
$\|\mathcal P_{\ker A^{\mathsf T}}y\|$ as $\alpha\to0$ to $\|y\|$ as $\alpha\to\infty$,
so target selection must respect both limits and the finite search interval.
The matrices used here have full row rank, placing the lower limit at zero;
every recorded target lies below the data norm and its root is bracketed.
These observations establish attainability for the reported fits. Both limits are
properties of the data space.
The unobservable isotropic gauge lies in the parameter space and is already removed from
the solved coordinates, so it cannot raise the lower limit.

At the small-parameter end, a flat GCV score scarcely distinguishes neighboring
parameters. The selected fits are near-interpolating and remain bounded in the
reported calculations; a lower-edge selection alone does not establish divergence.
Normal-equation relative residuals below $1.1\times10^{-14}$ check satisfaction of
the solved equations, rather than solution stability. Trace deficits, plateau
diagnostics and selected indices, including those for the constitutive-field study,
are recorded in the Supplementary Material.

What the present tests support is therefore an empirical statement rather than a ranking
of the two rules: the stripe improvement holds under both, and a selected parameter
should be reported together with its position in the search interval and the local
behavior of its criterion. Comparing how the two rules respond to forward-model error
would need an experiment that specifies that error. The transfer test here does not
supply one, because the four-roll observations retain the $n=95$ source bandwidth while
the inverse representation is $n=33$, which is a representation gap rather than an error
in the declared covariance.

\section{Conclusions}

Two calibrated optical components determine the deviatoric stress of a planar viscoelastic
flow pointwise, and velocity constrains how that stress varies in space through momentum
balance. The two channels are complementary rather than redundant, and that complementarity is
what supplies independent physical information across an optical gap.

The isotropic part of the stress is unobservable to both channels. The optical pair sees
only the deviatoric part by construction, and the divergence of an isotropic field is a
pure gradient, which the Leray projection annihilates. The joint operator is therefore
rank deficient by at least that number for any sampling, and by exactly that number
when the optical field is complete; we confirmed the count on three discretizations.
The finite, incomplete, aperture-averaged operators used here are rank deficient by
more than that, so the gauge is the part of the deficiency that no sampling removes. Any reconstruction target must be
stated as a quotient that removes this gauge; reporting an error on the full tensor would
otherwise charge the method for a component no measurement of this kind can supply.

With that target, velocity contributes substantially where the optical map is incomplete.
In paired tests with finite apertures and spatially correlated noise, stripe dropout gives
a mean whole-domain deviatoric error of $50.40\%$ from optics alone and $27.82\%$ jointly
at $3\%$ reference noise. The improvement persists when the velocity output centers in the same stripe are
removed as well, though the retained apertures still overlap it. It survives refinement of the inverse representation at fixed physical
sampling, and it persists for a stress field generated by integrating a finitely
extensible constitutive model rather than drawn from the reconstruction ensemble.

The regularization choice needs more care than the agreement of two standard
selection rules suggests. Here the targets are reachable: the observation matrices have full row rank, which
puts the attainable residual at zero, each target falls below the norm of its data,
and every reported discrepancy root is bracketed. Generalized cross-validation
reaches the lower edge of its grid in a substantial fraction of fits. A selected parameter should be reported with its position in the search
interval and the local behavior of its criterion; section~\ref{sec:discussion}
sets out both diagnostics.

The reconstruction results are stated for the prescribed synthetic ensembles, apertures and covariances, and transfer to an
instrument would require the measured joint covariance and spatial response in the same
observation model. The identifiability statement, by contrast, is structural: it depends on
the operators and not on the noise, and it will hold for any sampling of this measurement
pair.

\roles{%
\textbf{Zijian Liu:} Conceptualization, Methodology, Software, Formal
analysis, Investigation, Funding acquisition, Writing -- original draft,
Writing -- review and editing.
\textbf{Julian Olszewski:} Conceptualization, Methodology, Software, Formal
analysis, Investigation, Funding acquisition, Writing -- original draft,
Writing -- review and editing.
\textbf{Bruce I. Gaynes:} Conceptualization, Writing -- review and editing.
\textbf{Jie Xu:} Conceptualization, Methodology, Formal analysis, Investigation, Supervision,
Writing -- review and editing.%
}

\ack{%
The authors declare no conflicts of interest.
The authors used GPT-6 (OpenAI) and Claude Opus 5 (Anthropic) to assist with
literature search, coding, language, organization, and readability. The
authors reviewed and edited the resulting material as needed and take full
responsibility for the content of the publication.%
}

\funding{%
Z.L.\ acknowledges support from the Illinois Society for the Prevention of Blindness (ISPB).
J.O.\ acknowledges support from the University of Illinois Chicago
Chancellor's Undergraduate Research Award (CURA).
B.I.G.\ acknowledges support from the Richard A. Perritt, MD Charitable
Foundation.%
}

\data{%
The data that support the findings of this study are available upon reasonable
request from the authors.%
}

\begingroup\footnotesize\setlength{\bibsep}{0pt}\raggedright\interlinepenalty=10000

\endgroup
\end{document}